\documentclass[aps,prx,a4paper,oneside,12pt]{revtex4-1}

\usepackage[ngerman]{babel}
\usepackage[T1]{fontenc}
\usepackage[ansinew]{inputenc}
\usepackage{amsfonts}
\usepackage{amsmath}
\usepackage{amssymb}
\usepackage[colorlinks,linkcolor=blue,citecolor=blue,urlcolor=blue]{hyperref}
\usepackage{lmodern} 
\usepackage{graphicx} 
\usepackage{amsmath}
\usepackage{amsthm}
\usepackage{amsfonts}

\begin{document}

\title{Tunable High Harmonic Pulses  from Nanorings Swirled by Optical Vortices}

\author{J. W\"{a}tzel$^1$} 
\email{jonas.waetzel@physik.uni-halle.de}
\author{J. Berakdar$^1$}
\email{jamal.berakdar@physik.uni-halle.de}
\affiliation{$^1$Institute for Physics,  Martin-Luther-University Halle-Wittenberg, 06099 Halle, Germany}

\begin{abstract}
Irradiating  intercalated nanorings by  optical vortices   ignites  a   charge flow that
emits coherent  trains of high harmonic bursts with frequencies and time structures that are controllable  by the topological charge of the driving vortex beam. Similar to  synchrotron radiation, the polarization  of  emitted harmonics   is also selectable   by tuning to the appropriate  emission angle with respect to the ring plane. The nonequilibrium orbital magnetic moment  triggered  in  a ring tunnels quantum mechanically  to smaller and larger  rings  leading respectively to   high and low-frequency  harmonic generation.
 The frequencies of the emitted harmonics are  tunable by simply changing  the waist and/or the winding number of the optical vortex, without the need to increase the pulse intensity which  can  lead to material damage.
These findings follow from full-fledged quantum dynamic simulations for realistic material and laser parameters. The proposed setup is  non-destructive as only  short vortex pulses of moderate intensities   are needed, and  it offers a versatile tool for nanoscale optical and spectroscopic applications such as local,  single beam pump-probe experiments.
\end{abstract}
 
\maketitle

\section{Introduction}

\begin{figure}[t]
\centering
\includegraphics[width=12cm]{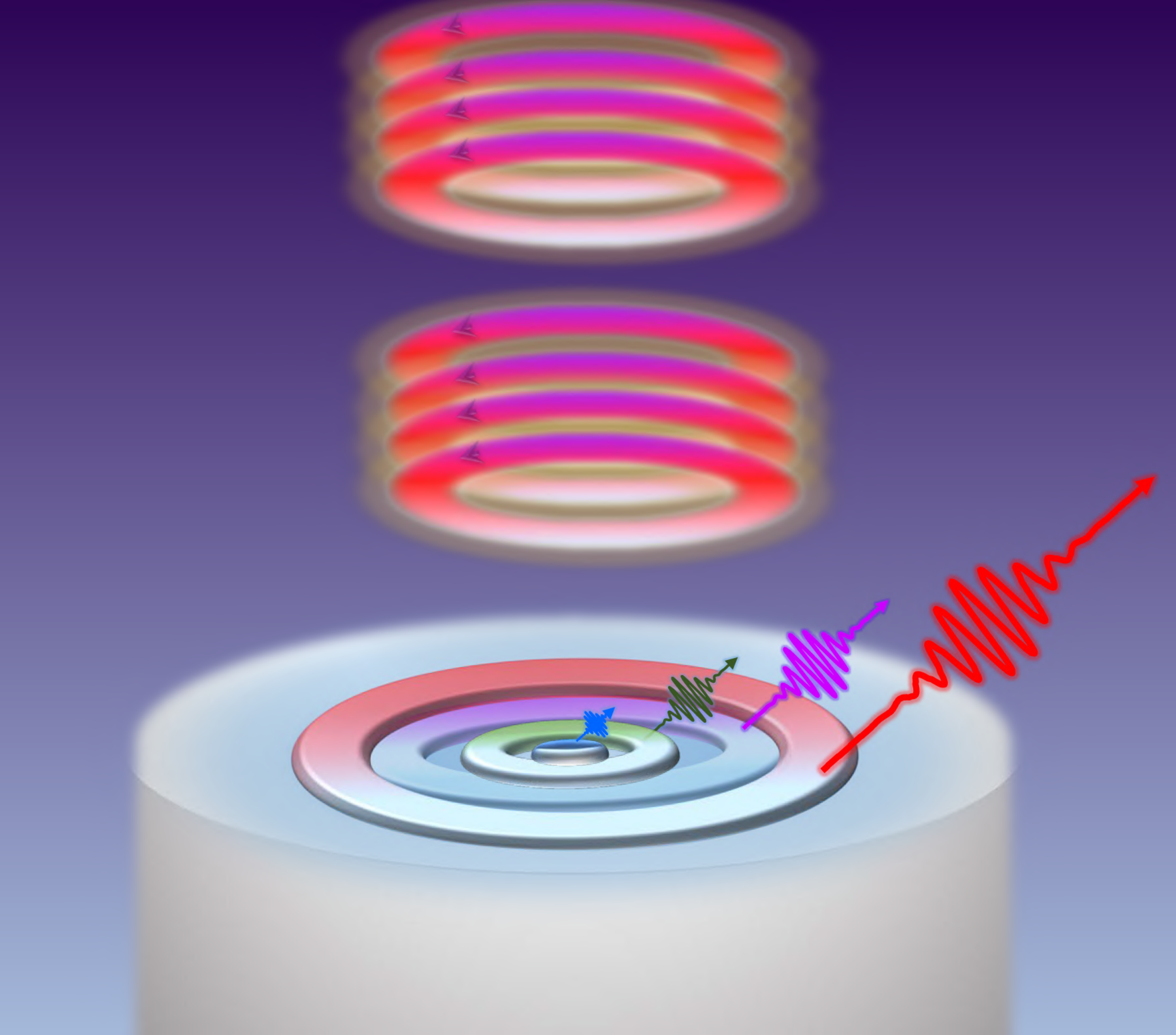}
\caption{Schematics of the considered setup. A focused  optical vortex beam transfers orbital angular momentum to a quantum ring with an amount related to the topological charge $m_{_{\rm OAM}}$ of the vortex triggering so a charge current loop around the ring. The photoexcited current tunnels then  successively to neighboring intercalated rings. The so generated non-equilibrium circulating charge distributions emit pulses of  photons with  specific frequencies and delayed by roughly  the inter-ring tunneling and rise-up times. The pulses are circularly polarized in the direction vertical to the ring planes and their frequencies can be tuned by the topological charge of the optical vortex at a fixed laser frequency and intensity. Changing the waist of the vortex focuses the beam onto a ring with a desired radius and allows for up or down conversion of the laser frequency.}
\label{fig:fig0}
\end{figure}

Radiation emission from accelerated  charge in synchrotron ring facilities has played a key role in the advancement of modern science. Envisioning a "nano synchroton"  with  charge distribution  looping around and radiating in nano rings  offer  a  nanoscale optical source with wide ranging applications.  Such rings can be deposited for instance  on a scanning tip enabling so a local  probing.
Two aspects are important. To reduce losses, phase coherent rings are appropriate in which case the charge acceleration and emission should be considered quantum mechanically.  To power the rings one may think of applying magnetic field pulses. Exceedingly larger magnetic fields are needed for smaller rings  however, entailing   high power consumption to produce these fields which  couple only weakly to charge.
We find, utilizing the electric field of an optical vortex, combined with a smart  engineering for charge confinement in form of  intercalated nanorings
yield  an elegant  and efficient "nano synchroton". As evidenced  below (cf. Fig.\,\ref{fig:fig0}), the emitted radiation due to the looping current  is simply controllable by the winding number of the vortex (also called the topological charge  $m_{_{\rm OAM}}$) at a fixed frequency and a moderate intensity of the vortex beam.  As for the size of this nano-quantum-synchrotron source,  the diffraction (Abbe) limit of the driving pulse sets a clear limitation; a ring  with a size well smaller than the optical wavelength lies in the "dark zone" of the vortex and is only very weakly perturbed (cf. Fig.\,\ref{fig:fig0}). This can be circumvented by letting the vortex-induced current tunnels to the smaller, appropriately engineered rings  generating so higher harmonics, as demonstrated below.  The Abbe diffraction limit poses no obstacle on the size of the rings and can thus be beaten by enclosing rings much smaller than the vortex waist. To increase the current and to tunnel to tinier rings emitting at higher frequencies, one should enhance the vortex winding number    $m_{_{\rm OAM}}$, not the intensity, to avoid material damage and make the concept feasible for solid state realization. In contrast, our method is not appropriate for the gas phase, due to the vast mismatch between the electronic orbital size and vortex waist. An establlished method for gas-phase harmonic generation  is based on bound electron tunneling, acceleration, and coherent re-scattering and/or recombination  in spatially homogeneous laser fields with intensties several orders in magnetiude higher than the vortex beam intensities \cite{Note1} used here \cite{3stepmodel1,3stepmodel2,BrabecRMP}.
As shown below the current triggered in the irradiated ring tunnels quantum mechanically  to the inner and outer rings generating non-equilibrium orbital  moments in the respective rings that emit at their individual characteristic frequencies.  Interestingly, the tunneling times between the rings  result in a natural time delay of the emitted harmonics from different rings (as schematically illustrated in  Fig.\,\ref{fig:fig0}). As these tunneling times can be controlled by the height and width of the barriers separating the rings (for example by an appropriate gate voltage), the time delay between the harmonics is also controllable. The proposed optical source is therefore useful  for local  pumping a sample, such as adsorbate or surface, by  radiation from one ring and probing the excitation after a time delay by  radiation from another ring, similarly as done in two-photon photoemission spectroscopy to access and trace the excited states dynamics \cite{2ppe1,2ppe2,2ppe3}.
Notably, the concept presented here is non-invasive as only moderate intensity vortex pulses are needed with a duration of a few optical cycles.
\\
In principle, one may attempt utilizing conventional  circular polarized Gaussian pulses, but optical vortices (also called beams carrying  orbital angular momentum (OAM)) offer more flexibility due to their specific spatial structure and the fact that the amount $m_{_{\rm OAM}}$ of the carried and transferable OAM can be increased, in principle without limit. As shown below the vortex OAM is intimately related to the strength of the generated charge current loop in the ring. \\
Generally, the total angular momentum of electromagnetic waves has a spin angular momentum (SAM) and an orbital angular momentum (OAM) component, both  were analyzed theoretically  long  ago \cite{rosenfeld1940on, belinfante1940current, tannoudji1989atoms, soper2008classical}.
Meanwhile,  light modes carrying OAM are  feasibel at a wide range of pulse parameters  \cite{allen1992orbital, beijersbergen1993astigmatic, beijersbergen1994helical, he1995direct, simpson1997mechanical, soskin1997topological, allen2003optical} and have been used for applications ranging from optical tweezers for microscale objects to electronics and life sciences, quantum information or optical telecommunications \cite{molina2007twisted, mair2001entanglement, barreiro2008beating, boyd2011quantum, padgett2011tweezers, furhapter2005spiral, woerdemann2009self, torres2011twisted, andrews2011structured, foo2005optical, he1995optical, wang2008creation, hell2007far}. OAM pulses  enabled also photomechanics  for moving, trapping and rotating microscopic objects \cite{allen2002introduction, barreiro2003generation, friese1998optical}, atoms, molecules \cite{romero2002quantum, al2000atomic, araoka2005interactions, watzel2016optical}, and Bose-Einstein condensates \cite{helmerson2011rotating} as well as a charge distribution \cite{watzel2016centrifugal}.\\
On the material side, advances in fabrication and patterning of micro and nano structures  produced a fascinating variety of nanobjects \cite{bruchhaus2017comparison}. Here we focus on phase coherent quantum ring structures which are a prototypical example for studying non-equilibrium dynamics \cite{kravtsov1993direct, chalaev2002aharonov, matos2004field, matos2004ultrafast, pershin2005laser, matos2005photoinduced, moskalenko2006revivals, moskalenko2008polarized, hinsche2009high, barth2006unidirectional, barth2007electric}. Vortex beam  may act with a torque  on the charge carriers \cite{friese1998optical, o2002intrinsic, simpson1997mechanical, gahagan1996optical, babiker1994light, andrews2012angular} leading for instance to photo-induced directed current amd photogalvanic effects with a strength controllable by  $m_{_{\rm OAM}}$ \cite{watzel2016centrifugal}.
To be specific, results and numerics are presented for semiconductor-based quantum rings irradiated by an optical vortex beam. As  made clear below, mathematically and by physical reasoning one concludes that
the scheme is a general nature, and the phenomena should be observable at different frequencies, ring sizes and materials compositions. Our particular focus is to explore theoretically the feasibility of the setup in Fig.\ref{fig:fig0} as a source for circular polarized trains of radiations adjustable by  $m_{_{\rm OAM}}$. Such a source might be useful for instance for spatially resolved pump-probe experiments. To this end we perform full numerical simulations for the charge dynamics in the coupled quantum rings irradiated by the OAM beam and employ the theory of a time-dependent spectrum to retrieve the emitted radiation from the nonequilibrium population \cite{eberly1977time, raymer1995ultrafast, moskalenko2008polarized, moskalenko2017charge}. The frequencies of the emission spectrum are controllable by $m_{_{\rm OAM}}$ and can be up- and down-converted by simply changing the waist of the OAM beam. The time delay between different emitted bursts is tunable by tweaking the tunneling barrier between the rings.
\section{Vortex-matter interaction}
The ring structures depicted in Fig.\,\ref{fig:fig0} have four  relevant length scales (with distinctly separated typical  energy scales of charge excitations), the atomic-sized lattice constant $a$, the $nm$  ring height  $\Delta h$ confinement, the tens of $nm$ ring width  $\Delta \rho_j$,  and  the  radius $\rho_j$  of the ring $j$. Typically $\rho_j$ is in the range of  100 nm to microns. One $\rho_j$ is chosen to be comparable with the pulse waist whose extension is bound by the diffraction limits. Other rings are excited by tunneling currents and their sizes may be way smaller.  Here the driving pulse frequency $\omega_x$ is tuned to be below the excitation frequency of the first quantum well state of the vertical confinement,  meaning $\omega_x<3\pi^2\hbar/(2m^* \Delta h^2)$ and $m^*$ is the effective mass of the GaAs heterostructure. Under these conditions, only quantum states, labelled respectively $m_{_0},\ n_{_0}$ due to finite size of $\rho_j$ and $\Delta \rho_j$ are affected by the laser pulse (a typical result is in  Fig.\,\ref{fig:fig1}a). Interestingly, the OAM transferred by the vortex lifts the conventional optical selection rules increasing so the number of accessible final states  that are within the spectral width of the pulse \cite{watzel2016centrifugal, watzel2016optical}. The laser intensity is kept moderate to avoid complications related to  multiphoton and other strongly nonlinear  processes which may lead to coupling to higher energy modes and strong  current relaxation heating the sample.  Within this outlined  setting we capture still all higher order effects in the charge-vortex interaction by propagating on a space-time grid the single particle states $\Psi_{n_{_0},m_{_0}}(x,y,t)$  in the presence of the vortex field according to (the ring is in the $x-y$ plane)
\begin{eqnarray}
i\hbar\partial_t\Psi_{n_{_0},m_{_0}}(x,y,t) & = &\left[-\frac{\hbar^2}{2m^*}(\partial_x^2+\partial_y^2) + \frac{ie\hbar}{2m^*}
\left(2\pmb{A}(x,y,t)\cdot\pmb{\nabla}+\pmb{\nabla}\cdot\pmb{A}(x,y,t)\right)\right. \nonumber \\
& &\left.+\frac{e^2}{2m^*}\pmb{A}(x,y,t)^2\right]\Psi_{n_{_0},m_{_0}}(x,y,t)
\label{eq:2D-TDSE}
\end{eqnarray}
where the charged particle mass $m^*$ stems from the effective mass approximation which can be applied because the energy dispersion $E(\pmb{k})$ in case of GaAS is approximately parabolic and isotropic for the states with few meV of energy around the $\Gamma$-point. Since we consider only moderate intensities of the vortex pulses, this approximation holds also for the minimal coupling Hamiltonian where the momentum is shifted by the vector potential $\pmb{A}(x,y,t)$ of the electromagnetic perturbation which has a moderate intensity.\\
In Eq.\,\eqref{eq:2D-TDSE} we use explicitly the single-particle picture which bases on neglecting Coulomb-interactions between the charge carriers. There are several reasons which legitimate this non-interacting particle picture: In the stationary case, i.e. $t\rightarrow-\infty$, it can be demonstrated that for a small number of electronic states the inclusion of correlation via Coulomb matrix elements in a quantum ring with no impurity simply shifts the non-interacting energy spectrum to higher energies \cite{pietilainen1995electron}. Therefore, we calculated numerical various Coulomb matrix elements $V_{abcd}=\langle\Psi_a\Psi_b|V(\pmb{r}-\pmb{r}')|\Psi_c\Psi_d\rangle$ where $V(\pmb{r}-\pmb{r}')=e^2/(4\pi\epsilon_0\epsilon_r|\pmb{r}-\pmb{r}'|)$ and can confirm that they are actually smaller than the kinetic Matrix elements. As a direct consequence on can deduce, that the correlation effects lead only to a small perturbation of the underlying electronic spectrum. But most importantly, the Coulomb matrix elements conserve the angular momentum, i.e. $m_a=m_c$ and $m_b=m_d$ \cite{zurita2002multipolar}. Since we consider optically induced transitions which change the internal angular momentum state, the Coulomb matrix elements between the states which are involved in the electric transition disappear. Consequently, correlation effects play a minor role for the qualitative description of the considered interband transitions. Thus, the parameters of the pulses are chosen so as to trigger intersubband excitations near the Fermi-level which alter the internal angular momentum state in which case the independent single-particle picture is viable \cite{tan1996electron, imry2002introduction, presilla1997nonlinear, chakraborty1994electron, matos2005photoinduced}.\\
The vortex vector potential satisfies the scalar Helmholtz equation in paraxial approximation \cite{allen1992orbital, beijersbergen1993astigmatic, beijersbergen1994helical, he1995direct, simpson1997mechanical, soskin1997topological, allen2003optical}. In the $x-y$ plane ($z=0$) in polar coordinates it reads
\begin{equation}
\pmb{A}(\rho,\varphi,t)={\rm Re}\left\{\hat{\epsilon}A_0f_{m_{_{\rm OAM}}}^p(\rho)\Omega(t)e^{i(m_{_{\rm OAM}}\varphi-\omega_xt)}\right\},
\end{equation}
where $\hat{\epsilon}$ is a polarization vector, $\varphi=\arctan[x,y], \rho=\sqrt{x^2+y^2},$ and $A_0$ is the field amplitude, and $\hbar\omega_x$ is the photon energy of the beam. The temporal envelope of the pulse is given by $\Omega(t)=\sin^2[\pi t/T_{\rm dur}]$ for $0<t<T_{\rm dur}$. In the case of charge carrier dynamics restricted to the $x-y$ plane, the radial distribution function is given by
\begin{equation}
f_{m_{_{\rm OAM}}}^p(\rho)=C_{\left|m_{_{\rm OAM}}\right|}^p L^{\left|m_{_{\rm OAM}}\right|}_p\left(\frac{2\rho}{w_0}\right)^{\left|m_{_{\rm OAM}}\right|}e^{-\frac{\rho^2}{w_0^2}}.
\end{equation}
Here, $C_{\left|m_{_{\rm OAM}}\right|}^p$ is the normalization constant, $L^{\left|m_{_{\rm OAM}}\right|}_p\left(x\right)$ are the associated Laguerre-Gaussian polynomials. $p$ is the radial node index and $w_0$ is the waist of the laser spot. For clarity we focus on $p=0$, meaning a donut shape radial intensity distribution.\\
Technically, we assume the ring structure to be imprinted on a GaAs-AlGaAs-based two dimensional electron gas. The radial confinement potential of the quantum ring is given by $V(\rho)=a_1/\rho^2 + a_2\rho^2 - V_0$ \cite{tan1996electron}  with  $V_0=2\sqrt{a_1a_2}$. The average ring radius is $\rho_0=(a_1/a_2)^{1/4}$, while the average ring width for a given Fermi energy $E_F$ is $\Delta\rho\approx\sqrt{ 8E_F/m^*w_0^2}$. Here, the oscillator frequency is $\omega_0=\sqrt{8a_2/m^*}$. For a radius $\rho$ near the average radius $\rho_0$, the confinement potential is parabolic $V(\rho)\approx\frac{1}{2}m^*\omega_0^2( \rho-\rho_0)^2$. A quantum dot is achieved for $a_1=0$. The stationary states of this quantum ring are  labeled as
\begin{equation}
E_{n_{_0},m_{_0}}=\left(n_0+\frac{1}{2}+\frac{1}{2}\sqrt{m_0^2+\frac{2m^*a_1}{\hbar^2}} \right)\hbar\omega_0 - \frac{m^*}{4}\omega_0^2\rho_0^2
\end{equation}
 $n_0$ and $m_0$ are  radial  and  angular quantum numbers. The minima of all subbands are at $m_0=0$ and the subband energy spectra are symmetric with respect to $m_0=0$, and due to time-reversal symmetry they  are degenerated with respect to the clock-wise and anti-clock-wise angular motion, i.e. $E_{n_{_0},m_{_0}}=E_{n_{_0},-m_{_0}}$ (currentless ground states).
\section{Emission spectrum}
Having triggered a circulating non-equilibrium charge distribution in the ring we are interested in the emitted radiation in the far field, meaning at a point
with a distance $r=|\pmb{r}|$ that is much larger than the diameter of the largest ring.
The intensity per spherical solid angle $\Omega(\vartheta,\varphi)$ of the far-field radiation  is given by $\frac{dI}{d\Omega}=\frac{8}{3}c\epsilon_0r^2\left|\langle \pmb{E}(\pmb{r},t)\rangle\right|^2$ \cite{jackson1999classical}. Here, $c$ is the speed of light, $\epsilon_0$ the vacuum dielectric constant, $\pmb{E}(\pmb{r},t)$ the electric field of the emitted radiation, and $\langle \cdot\cdot\cdot\rangle$ stands for the expectation value. The theory of the time-dependent spectrum yields the filtered radiant intensity $I(\omega,\Omega,t)$ in the direction $\pmb{n}\|\pmb{r}$ depending on the frequency $\omega$, and the detection time $t$ \cite{eberly1977time, raymer1995ultrafast}. The positive-frequency part $\pmb{E}^{(+)}(t)$ and the negative-frequency part $\pmb{E}^{(-)}(t)$ of the electric field component $\pmb{E}(t)$ are defined by
\begin{equation}
\pmb{E}^{(+)}(t)=\left(\pmb{E}^{(-)}(t)\right)^*=\frac{1}{2\pi}\int{\rm d}\omega\,\pmb{\widetilde{E}} (\omega)\Theta(\omega)e^{-i\omega t}
\end{equation}
where $\Theta(\omega)$ is the unit step function and $\pmb{\widetilde{E}}(\omega) = \int{\rm d}t\,\exp(i\omega t)\pmb{E}(t)$. For an observer at a distance $r$ the detected physical radiant intensity spectrum of radiation in direction $\pmb{n}_\alpha$ is given by the truncated fourier transforms \cite{eberly1977time,courtens1977time,renaud1977nonstationary,brenner1982time}:
\begin{equation}
\begin{split}
\frac{d^2I_\alpha}{d\omega d\Omega}=&\frac{8}{3}c\epsilon_0r^2
\int_{-\infty}^{\infty}{\rm d}t'\int_{-\infty}^{\infty}{\rm d}t''G(t-t')G(t-t'')
e^{i\omega(t''-t')} \left(\pmb{n}_\alpha\cdot\pmb{E}^{(+)}(t')\right)
\left(\pmb{n}^*_\alpha\cdot\pmb{E}^{(-)}(t'')\right).
\end{split}
\label{eq:PowerSpectrumI}
\end{equation}
Here, the detection window function is defined as $G(t)=\left(\frac{2} {\pi}\right)^{1/4} \frac{1}{\sqrt{\Delta T}}e^{-t^2/\Delta T^2}$ where $\Delta T$ mimics the detection time interval. 
By introducing the positive- and negative-frequency part of the filtered (by the detector) electric field
\begin{equation}
\pmb{\overline{E}}^{(+)}(\omega,t)=\left(\pmb{\overline{E}}^{(-)}(\omega,t)\right)^*=\int{\rm d}t'\,\pmb{E}^{(+)} (t')G(t-t')e^{-i\omega t'}
\label{eq:FilteredEfield}
\end{equation}
the time-dependent physical spectrum of the radiation can be found as
\begin{equation}
\begin{split}
\frac{d^2I_\alpha}{d\omega d\Omega}=&\frac{8}{3}c\epsilon_0r^2
\left(\pmb{n}_\alpha\cdot\pmb{\overline{E}}^{(+)}(\omega,t)\right)
\left(\pmb{n}^*_\alpha\cdot\pmb{\overline{E}}^{(-)}(\omega,t)\right)
\end{split}
\label{eq:PowerSpectrumII}
\end{equation}
which is strictly positive for all frequencies $\omega$ and times $t$.\\
The coherent electric field part of the emitted radiation  of the driven ring is \cite{jackson1999classical,landau2000classical}
\begin{equation}
\pmb{E}(\pmb{r},t)=\frac{1}{4\pi\epsilon_0c^2r}\left\{\pmb{n}\times\left[\pmb{n}\times\ddot{\pmb{\mu}}(t-t_d)\right] + \frac{1}{6c}\pmb{n}\times\dddot{\pmb{D}}(t-t_d) + \frac{1}{c}\pmb{n}\times\ddot{\pmb{m}}(t-t_d)\right\},
\label{eq:ElectricField}
\end{equation}
where $\pmb{\mu}(t)$ is the dynamical electric dipole moment, $\pmb{m}(t)$ the dynamical magnetic dipole moment and $\pmb{D}(t)$ the quadrupole moment of the driven quantum ring. The delay time is determined by the distance between the ring and the observer and is given by $t_d=r/c$.\\
In the following, the quantum ring structure  is  in the $x-y$ plane while we detect the emitted radiation in $z$-direction, i.e. $\pmb{n}=\hat{e}_z$. Since the magnetic moment $\pmb{m}(t)$ also points in this direction, only the electric part of the radiation  contribute to the whole emission power. The vector of electric quadrupole moment is defined by the tensor product $\pmb{D}=\sum_{\beta}D_{\alpha\beta}n_{\beta}\pmb{n}$ which depends in both the magnitude and direction, on the direction to the point of observation $\pmb{r}$. The quadrupole tensor is given as $D_{\alpha\beta}=\int{\rm d}\pmb{r}\,\rho(\pmb{r})\cdot \left(3r_\alpha r_\beta - r^2\delta_{\alpha\beta}\right)$ where $r_\alpha=x,y,z$. Consequently, the quadrupole vector $\pmb{D}=(D_{xz},D_{yz},D_{zz})^T$. Since the quantum ring is located in the $x-y$ plane we infer that $D_{xz}=D_{yz}=0$. Therefore, according to Eq.\,\eqref{eq:ElectricField} no quadrupole radiation signal is expected  in the direction $\pmb{n}$ to the observer $\pmb{r}$ because also $D_{zz}$  time-averages to zero.\\
The radiation polarization is classified by the Stokes parameter $S_0$, $S_1$, $S_2$ and $S_3$ \cite{mcmaster1954polarization}. The parameter $S_0$ indicates the intensity while $S_1$ and $S_2$ quantify the linear polarization. $S_3$ signifies   circularly polarized radiation. The degree of polarization is given by $p=\sqrt{S_1^2+S_2^2+S_3^2}/S_0$. In order to find the mathematical expressions for the Stokes parameter we use the following polarization vectors which are perpendicular to $\pmb{n}$: $e_x$ and $e_y$ in Cartesian basis, $e_{\pm45^\circ}=\frac{1}{\sqrt{2}}(e_x\pm e_y)$ in a Cartesian basis rotated by $45^\circ$, and $e_{\pm}=\frac{1}{\sqrt{2}}(e_x\pm ie_y)$ which describes the circular basis.\\
Considering Eq.\,\eqref{eq:PowerSpectrumII} the Stokes parameters read
\begin{equation}
S_{1(2,3)}(\omega,t,\Omega)=\frac{d^2 I_{x(45^\circ,+)}}{d\omega d\Omega} - \frac{d^2 I_{y(-45^\circ,-)}}{d\omega d\Omega}
\end{equation}
while the intensity is characterized explicitly by
\begin{equation}
S_{0}(\omega,t,\Omega)=\frac{d^2 I_{x}}{d\omega d\Omega} + \frac{d^2 I_{y}}{d\omega d\Omega}.
\end{equation}
At the observer $\pmb{r}$ the Stokes parameters describing the linear polarization can be computed according
\begin{equation}
S^z_1(\omega,t)=\frac{1}{6\pi^2\epsilon_0 c^3}\left(\left|\ddot{\mu}_x^{(+)}(\omega,t)\right|^2 - \left|\ddot{\mu}_y^{(+)}(\omega,t)\right|^2\right)
\end{equation}
and
\begin{equation}
S^z_2(\omega,t)=\frac{1}{6\pi^2\epsilon_0 c^3}{\rm Re}\left\{\ddot{\mu}_x^{(-)}(\omega,t)\ddot{\mu}_y^{(+)}(\omega,t) + \ddot{\mu}_x^{(+)}(\omega,t)\ddot{\mu}_y^{(-)}(\omega,t)\right\}
\label{eq:Stoke12}
\end{equation}
while the circularly polarization reads
\begin{equation}
S^z_3(\omega,t)=\frac{1}{6\pi^2\epsilon_0 c^3}{\rm Im}\left\{\ddot{\mu}_x^{(-)}(\omega,t)\ddot{\mu}_y^{(+)}(\omega,t) - \ddot{\mu}_x^{(+)}(\omega,t)\ddot{\mu}_y^{(-)}(\omega,t)\right\}.
\label{eq:Stoke3}
\end{equation}
The total emitted power follows as
\begin{equation}
S^z_0(\omega,t)=\frac{1}{6\pi^2\epsilon_0 c^3}\left(\left|\ddot{\mu}_x^{(+)}(\omega,t)\right|^2 + \left|\ddot{\mu}_y^{(+)}(\omega,t)\right|^2\right).
\label{eq:Stoke0}
\end{equation}
Fourier transforms are  computed  as  in Eq.\,\eqref{eq:FilteredEfield}.\\
In a related approach the temporal evolution of the spectrum of the emitted dipolar radiation can be obtained through a wavelet transform \cite{chui2016introduction,daubechies1992ten} of the acceleration as the second temporal derivative of the dipole moment $\pmb{\mu}(t)$. The time evolution of the spectrum can be found by a Gabor-type transformation $\mathcal{A}_w(t_0,\omega)=\int_{-\infty}^{\infty}{\rm d}t\,w_{t_0,\omega}(t)A(t)$ where the kernel function $w_{t_0,\omega}(t)$ has a variable width in a way that the number of oscillations is constant. Consequently, the shape of the analysing wavelet is unchanged. The main difference to the conventional Gabor transformation $\mathcal{A}_g(t_0,\omega)$ as the most used windowed Fourier transform lies in the Gebor kernel function $g_{t_0,\omega}(t)$ which has a fixed width and therefore, it could only be used when the frequency of the signal is neither too low or to high. We checked the emitted power spectrum obtained from Eq.\,\eqref{eq:Stoke0} with the wavelet transform methods and obtained similar radiation characteristics by using a Morlet kernel function \cite{de1996wavelet}.

\subsection{Pushing up  the emission frequency via  orbital angular momentum transfer}

\begin{figure}[t]
\centering
\includegraphics[width=12cm]{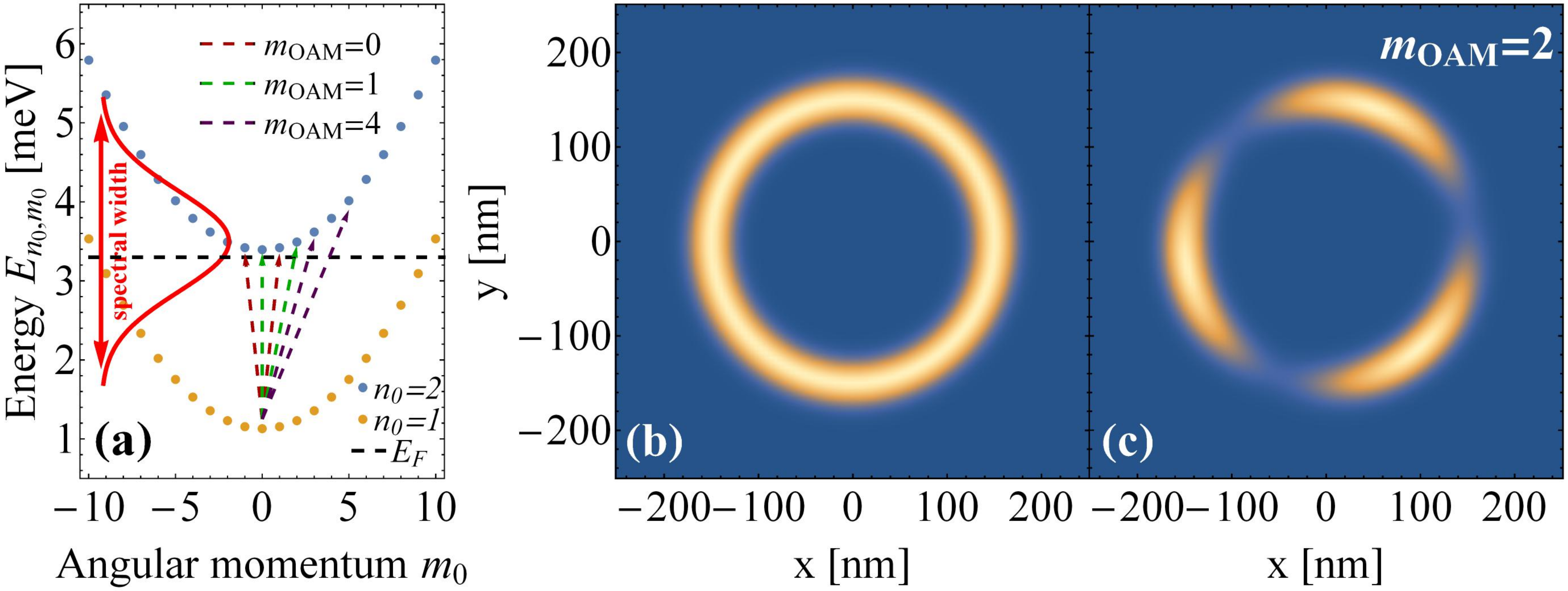}
\caption{(a) Energy spectrum of the quantum ring and transition scheme in dependence on the topological charge of the applied optical vortex beam. The fermi energy $E_F$ is marked by the dashed horizontal line. The red curve illustrates the spectral width of the employed two-cycle pulse. (b) The initial density of the occupied states $(t=0)$ corresponding to the ring structure. (c) snapshot of the excited population  $\tilde \rho(t)$ during the interaction with an optical vortex pulse with a topological charge $m_{_{\rm OAM}}=2$.}
\label{fig:fig1}
\end{figure}

\begin{figure}[t]
\centering
\includegraphics[width=12cm]{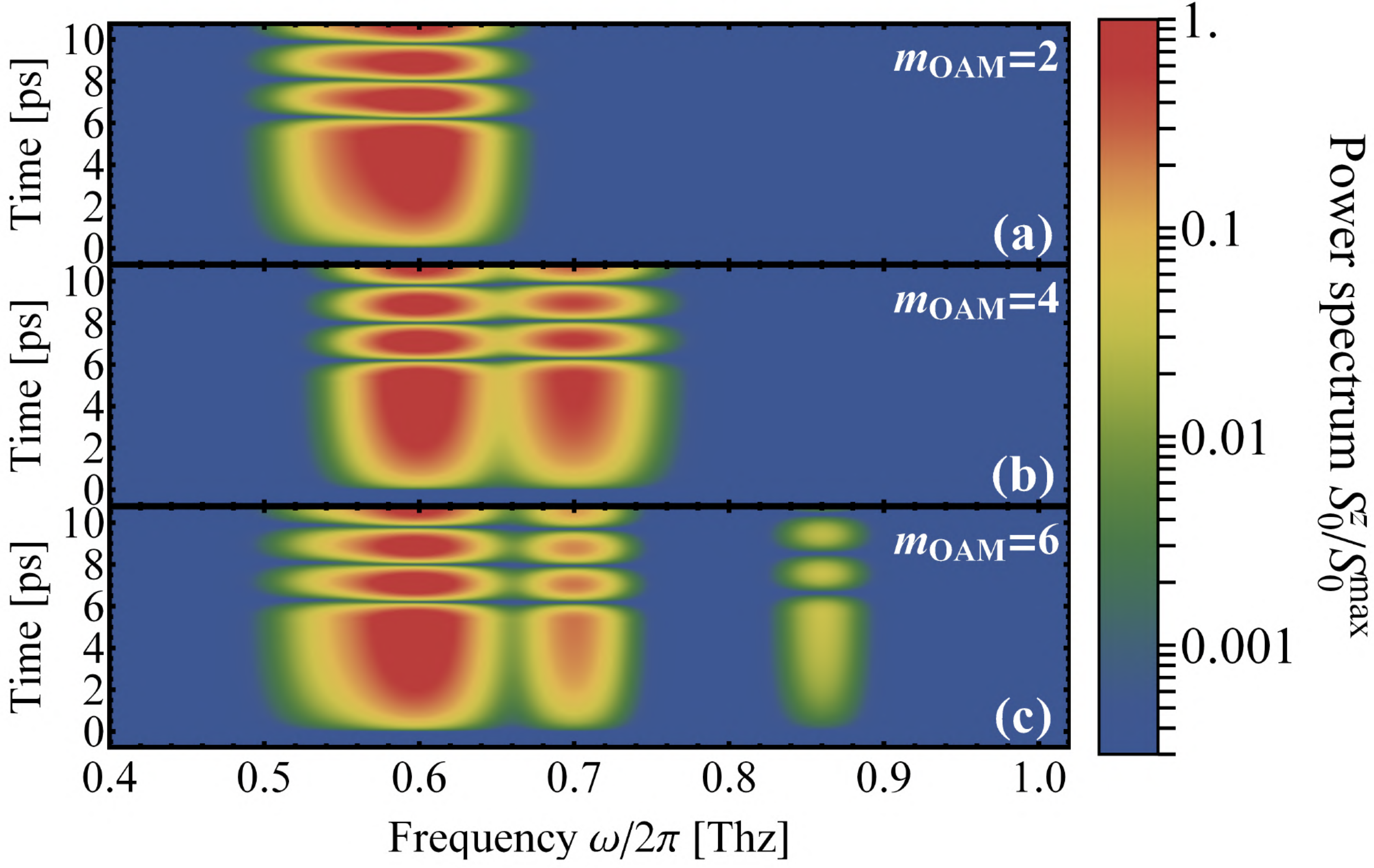}
\caption{(a) Time dependent emission spectrum for a two cycle optical vortex beam with a photon energy $\hbar\omega_x=2.5$\,meV. The topological charge of the OAM laser pulse is $m_{_{\rm OAM}}=2$. (b) the same as in (a) for a topological charge $m_{_{\rm OAM}}=4$. (c) the same as in (a) for a topological charge $m_{_{\rm OAM}}=6$.}
\label{fig:fig2}
\end{figure}

 Fig.\,\ref{fig:fig1}(a) shows the spectrum for a quantum ring with a radius $\rho_0=150$\,nm and an effective width of $\Delta\rho=40$\,nm.
 The ring is irradiated with a two-cycle long, linearly polarized  (along $x$) OAM pulse with a photon energy $\hbar\omega_x=2.5$\,meV which is
resonant with transitions from the first  to the second radial subband.   The Fermi level $E_F=3.3$\,meV (dashed horizontal line in Fig.\,\ref{fig:fig1}(a))
is set by the particle number which can be controlled for instance  by an appropriate gating.
The spectral width of the pulse is 3\,meV. The key advantage of vortex pulses is that the dynamics can be tuned by simply changing  the vortex topological charge
while keeping the intensity fixed (avoiding so multiphoton and other higher-order processes) and also the energy band width (and keeping so in resonance with the relevant states).
The peak intensity is chosen as  $10^{10}$\,W/cm$^2$. The laser spot is focused vertically  on the rings (cf. Fig.\ref{fig:fig0}).\\
For what follows it is crucial to recall that we are dealing with appropriately doped systems such
that the relevant  dynamic takes place in the subbands of the conduction band (in contrast to valence band excitations which are not discussed nor relevant here).
The kinetics of the   photoexcited population $\tilde \rho(t)$  depends on  the individual charge densities $\rho_{n_{_0},m_{_0}}$ corresponding to the wave functions $\Psi_{n_{_0},m_{_0}}(x,y,t)$ as yielded by the time-dependent Schr\"odinger equation  Eq.\,\eqref{eq:2D-TDSE} at the time $t$ for the states starting from the stationary states labeled $n_{_0},m_{_0}$
\begin{equation}
\tilde \rho(t)=\sum_{n_{_0},m_{_0}}f_{n_{_0},m_{_0}}(t)\left|\Psi_{n_{_0},m_{_0}}(x,y,t)\right|^2.
\end{equation}
The weights $f_{n_{_0},m_{_0}}(t)$ are the non-equilibrium distribution functions which can be evaluated by solving the Boltzmann equation within the relaxation time approximation \cite{ziman1972principles,matos2005photoinduced}:
\begin{equation}
\frac{\partial f_{n_{_0},m_{_0}}}{\partial t}=-\frac{f_{n_{_0},m_{_0}}(t)-f^0_{n_{_0},m_{_0}}(E_F)}{\tau_{\rm rel}}
\end{equation}
where $f^0_{n_{_0},m_{_0}}(E_F)=1/\left[1+\exp((E_{n_{_0},m_{_0}} - E_F)/k_BT)\right]$ is the Fermi-Dirac equilibrium distribution for a given temperature.
This simple description of the kinetics is reasonable as the highest excited state is still around the Fermi level. Within this picture, relaxation processes such as  electron-phonon scattering, electron-electron scattering, or simultaneous scattering by impurities and phonons are modelled  by a single (average) quantity, the relaxation time $\tau_{\rm rel}$. \\
To determine an appropriate value for $\tau_{\rm rel}$ we have to identify the relevant relaxation processes. First of all, we consider the multiple quantum ring system to be free from impurities. Second, Pauli blocking and the fundamental energy conservation limit effectively the redistribution of the charge carriers due to electron-electron collisions \cite{chakraborty1994electron}. Furthermore, in the case of GaAs the optical phonon energy is above 30\,meV \cite{grundmann1995inas} and much larger than the energy gap in the considered electronic structure [cf. Fig.\,\ref{fig:fig2}(a)]. We are considering optically induced transitions around the Fermi level $E_F$ and therefore, optical phonons will be not addressed. Rather, the relaxation process in our case and in the range of considered intensities are driven by scattering from incoherent phonons  \cite{moskalenko2017charge}. Time resolved theoretical and experimental studies of the intersubband relaxation processes in GaAs-AlGaAs quantum wells below the optical phonon energy revealed relaxation time scales between $10\,ps$ and $200\,ps$ \cite{murdin1996time}. We assume here a relaxation time $\tau_{\rm rel}=25\,ps$.\\
In Fig.\,\ref{fig:fig1}(b) the  population $\tilde \rho(t=0)$ right before the electromagnetic perturbation is shown.
It reveals, as expected an angularly homogeneous radial density distribution which is symmetric with respect to $\rho_0$. In the following we consider the ring perturbed by the focussed vortex beam radiation  causing dynamical changes in the density distribution. A typical example is shown in Fig.\,\ref{fig:fig1}(c) depicting the non-equilibrium population  $\tilde \rho(t)$ of the quantum ring during the interaction with the laser beam with a topological charge $m_{_{\rm OAM}}=2$. We identify   three nodal structures  related to the winding number \cite{watzel2016centrifugal}.\\
The variation of the topological charge enables transitions to final states with different magnetic quantum numbers $m_0$.
This is understandable by considering a perturbative picture \cite{watzel2016optical}: Crucial  for obtaining the optical  selection rules
 are the matrix elements $\langle\Psi_{n_{_0}',m_{_0}'}|H_{\rm int}(t)|\Psi_{n_{_0},m_{_0}}\rangle$ where $|\Psi_{n_{_0},m_{_0}}\rangle$ are the unperturbed ring eigenstates   and $H_{\rm int}(t)=\frac{ie\hbar}{2m^*}
\left(2\pmb{A}(x,y,t)\cdot\pmb{\nabla}+\pmb{\nabla}\cdot\pmb{A}(x,y,t)\right)
+\frac{e^2}{2m^*}\pmb{A^2}(x,y,t)$ is the  perturbation due to the optical vortex pulse.  For low intensities  $|\pmb{A}(x,y,t)|^2\approx0$ and the perturbative treatment for the linearly polarized vortex beam radiation leads to integrals in the form
\begin{equation}
\int_0^{2\pi}{\rm d}\varphi e^{-im_0'\varphi}\cos\varphi e^{im_{_{\rm OAM}}\varphi}e^{im_{_0}\varphi},
\label{eq:selection}
\end{equation}
from which  we deduce  the selection rules $m_0'=m_0+m_{_{\rm OAM}}\pm1$. For the special case $m_{_{\rm OAM}}=0$ resembling a linearly polarized Gaussian beam we infer  $m_0'=\pm m_0$ and therefore, no ring current is photo-induced \cite{watzel2016optical}. The possible transitions from the first radial band into the second in dependence on the topological charge $m_{_{\rm OAM}}$ are indicated in the transition scheme in Fig.\,\ref{fig:fig1}(a). From the matrix elements it can be deduced that the number of nodal structures in the density distribution of the ring [cf.\,Fig.\,\ref{fig:fig1}(c)] is given by $m_{_{\rm OAM}}+1$ \cite{watzel2016centrifugal}.\\
In Fig.\,\ref{fig:fig2} the time-dependent emission spectra are shown evidencing the dependence of the emitted frequency on the vortex winding number. This behavior we relate to the different optically induced (non-vertical) transitions between the first and second radial band. Another interesting aspect is that we find only emission in case of a even topological charge which leads to an odd number of nodal structures [cf.\,\ref{fig:fig1}(c)] \cite{watzel2016centrifugal}. This can be explained by a static picture. In the case of an even number of nodal structures we find always the same amount of charge density in every quadrant in the coordinate system relative to the center of the ring. Therefore, the dipole moments in the $x$- and $y$-direction equal to zero. This argument applies to the dynamical case where the nodal structures rotate with a round-trip frequency  depending on the parameter of the laser beam.
The employed OAM laser pulse has a photon energy $\hbar\omega_x=2.5$\,meV which means the maximum of the spectral width is centered on the transition $n_0=1,m_0=0\rightarrow n_0=2,m_0=3$ [cf.\,\ref{fig:fig1}(a)]. Therefore, we find a maximal emission signal when using a topological charge with $m_{_{\rm OAM}}=2$. The duration of the two-cycle pulse amounts  to $T_{\rm dur}=3.3$\,ps. The corresponding time-dependent spectrum $S_0^z(\omega,t)$ is shown in Fig.\,\ref{fig:fig2}(a). It exhibits repeated coherent radiation bursts centered around a frequency which corresponds to a photon energy of 2.47\,meV. The peaks decay in time due to  relaxation processes (relaxation time $\tau_{\rm rel}=25$\,ps). The repeated bursts are a consequence of the revivals of the charge polarization dynamics \cite{moskalenko2006revivals}.
Increasing  the topological charge to $m_{_{\rm OAM}}=4$ and $m_{_{\rm OAM}}=6$ results in additional emission signals at higher frequencies which we also expect on the basis of   the optical selection rules dependence on the topological charge. The intensity of the peaks reflects the spectral characteristics of the laser pulse. Therefore, in all cases the main maximum signal can be found around the central frequency $\omega_x$. In general, the number of peaks and the  frequency of the emission spectrum depend strongly on the topological charge $m_{_{\rm OAM}}$ as well as on the spectral width of the applied laser field.  For  a fixed frequency one has to consider the role the temporal  pulse length $T_{\rm dur}$ which sets  the spectral width \textbf{(see Appendix A)}.

\subsection{Up-conversion of emission frequency and tunneling of orbital magnetic moments}

\begin{figure}[t]
\centering
\includegraphics[width=12cm]{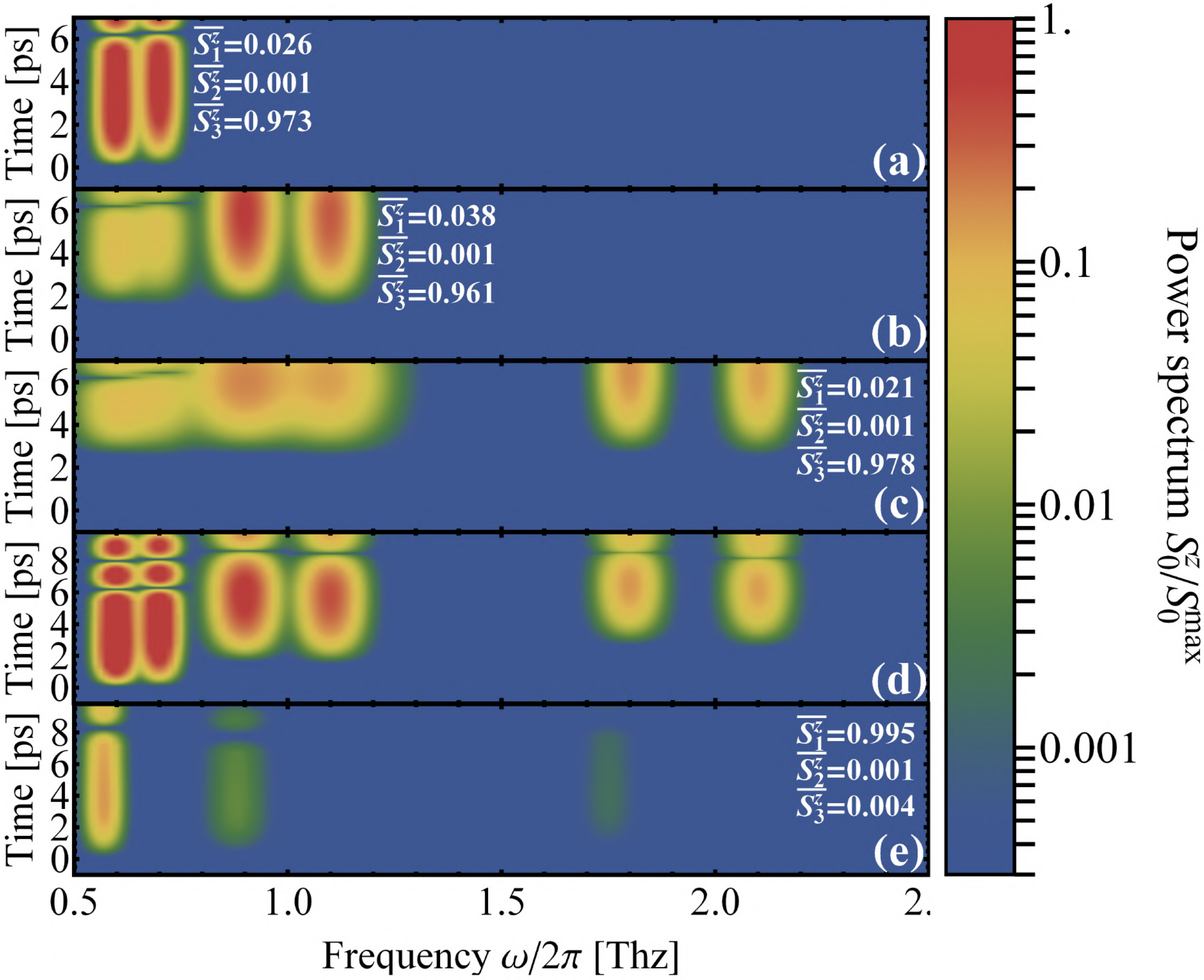}
\caption{Individual time-dependent emission spectra for the rings with (a) radius $\rho_1=150$\,nm, (b) $\rho_2=100$\,nm and (c) $\rho_3=65$\,nm. {The largest ring is irradiated by a focussed optical vortex beam with a topological charge $m_{_{\rm OAM}}=4$}. The spectra are normalized to the global maximal value. Therefore, the brightest color identifies the maximal intensity of the emission of the considered multiple ring-system. The total spectrum of the whole multiple-ring structure is shown in panel (d). {The spectrum in panel (e) shows the emission characteristics for a linearly polarized Gaussian beam with the  number of photons as for the vortex beam.}}
\label{fig:fig4}
\end{figure}

Considering multiple rings with different radii $\rho_1$, $\rho_2$ and $\rho_3$ coupled to each other we focus the OAM beam  on only one of the rings, the "perfect" vortex method is used \cite{jabir2016generation}. This new class of optical vortex beams is characterized by radii independent on the topological charge. A simple technique to generate such a beam is the Fourier transformation of the Bessel-Gauss beam with Fourier lenses of different focal lengths. The vector potential corresponding to the "perfect" vortex beam in the $x-y$ plane is
\begin{equation}
\pmb{A}(\rho,\varphi,t)={\rm Re}\left\{\hat{\epsilon}A_0e^{-\frac{(\rho-\rho_r)^2}{w_0^2}}
\Omega(t)e^{i(m_{_{\rm OAM}}\varphi-\omega_xt)}\right\}
\label{eq:perfectOAM}
\end{equation}
where $\rho_r$ and $w_0$ are the radius and annular width of the laser spot.
A schematics of the setup  is shown in Fig.\,\ref{fig:fig0}.  We conduct calculations for three intercalated  rings with the radii  $\rho_1=150$\,nm, $\rho_2=100$\,nm and $\rho_3=50$\,nm.  The effective widths are $\Delta \rho_1=\Delta \rho_2=\Delta \rho_3=40$\,nm, meaning that the width of the effective tunneling barriers between the rings is 10\,nm. Note, the largest quantum ring resembles the system and its characteristics used in Section A.
The  laser beam is focussed on the largest of the rings  while the others remain practically  unaffected. The pulse parameters are $\rho_r=150$\,nm and $w_0=10$\,nm with an intensity  at $10^{10}$\,W/cm$^2$.  Clearly, any  charge carrier dynamics within the "dark" rings is due to tunneling effects. We define the individual dipole moments corresponding to the specific rings $i$ as
\begin{equation}
\mu_x^i(t)=\sum_{n_{_0},m_{_0}}\int_{0}^{2\pi}{\rm d}\varphi\int_{\rho_i-\Delta \rho_i/2}^{\rho_i+\Delta \rho_i/2}{\rm d}\rho\,\rho^2\cos{\varphi}\left|\Psi_{n_{_0},m_{_0}}(\rho,\varphi,t)\right|^2
\label{eq:dipx}
\end{equation}
and ($\rho_i$ and $\Delta \rho_i$ are the radius and the effective width of the  ring $i$ )
\begin{equation}
\mu_y^i(t)=\sum_{n_{_0},m_{_0}}\int_{0}^{2\pi}{\rm d}\varphi\int_{\rho_i-\Delta \rho_i/2}^{\rho_i+\Delta \rho_i/2}{\rm d}\rho\,\rho^2\sin{\varphi}\left|\Psi_{n_{_0},m_{_0}}(\rho,\varphi,t)\right|^2.
\label{eq:dipy}
\end{equation}
  The corresponding power spectra follows  from Eq.\,\eqref{eq:Stoke0}.
In Fig.\,\ref{fig:fig4} the time-dependent emission spectra $S_0^{z,i}(\omega,t)$ corresponding to the individual rings $i$ as follows from the time-dependent dipole moments within  the respective domains $[\rho_i-\Delta \rho_i/2,\rho_i+\Delta \rho_i/2]$. The topological charge of the employed laser field is $m_{_{\rm OAM}}=4$.
The first emission spectrum Fig.\,\ref{fig:fig4}(a) reflects  the behavior observed  in Fig.\,\ref{fig:fig2}(b) very well. We find two different peaks around $\omega/2\pi=0.6$\,THz corresponding to the  optically accessible  main transitions. Furthermore, the depicted ring-resolved Stokes parameters $\overline{S^z_i}=S^z_i/S^z_0$ ($i=1,2,3$) reveal the polarization state of the emitted radiation. Here, the emission signal is largely dominated by circular polarization signifying the directed circular charge motion in the ring.\\
Two additional peaks occur when studying the emission spectrum of the second quantum ring in Fig.\,\ref{fig:fig4}(b). They are the result of tunneling processes from the larger ring into the smaller. At the same time, the rotational dynamics of the charge carriers is translated into the smaller quantum structure under conversation of the angular momentum, which amounts to a tunneling of the magnetic orbital moment. A smaller ring radius implies  a larger round-trip frequency of the looping charge carriers. However, the emission spectrum shows also signals at the same frequencies as in the spectrum corresponding to the  largest ring [cf. Fig.\,\ref{fig:fig4}(a)]. The tunneling probability and the effective tunneling time are strongly dependent on  the effective tunneling barrier which is also indicated by the emission spectrum in Fig.\,\ref{fig:fig4}(b). The intensities of the various peaks are weaker and they occur at later times. The displayed ring-resolved Stokes parameters correspond  to the frequency $\omega/2\pi=1.1$\,THz and  evidence the  circular polarization of the emitted radiation, which in turn signifies the tunneling of the orbital magnetic moment.\\
Considering the emission spectrum of the third smallest ring we observe same behavior, yet at higher frequencies (Fig.\,\ref{fig:fig4}(c)). Two
 additional signals occur at much higher frequencies with much weaker intensities. These structures can be traced back to a  tunneling process between the second and the third quantum ring. Furthermore, it causes a more pronounced retardation of the signals in time in comparison to the spectra corresponding to the first and the second ring. The maximum intensity of the emission signal can be found in the emission spectrum of the first ring.  The high harmonic  signals are circular polarized, as evidenced  by the displayed ring-resolved Stokes parameter $\overline{S^z_i}=S^z_i/S^z_0$.\\
In the panel (d) in Fig.\,\ref{fig:fig4} the emission spectrum of the whole multiple-ring spectrum is shown. It can be obtained by using the whole integration domain $[\rho_1-\Delta \rho_1/2,\rho_3+\Delta \rho_3/2]$ in Eq.\,\ref{eq:dipx} and Eq.\,\ref{eq:dipy}. It reveals the additional emerging signals at higher frequencies and later times, as already shown in the individual panels (a)-(c). Furthermore as already described above, these high harmonic signals have weaker intensities due to the tunneling processes from the large into the smaller rings. Clearly, attaching  additional rings with smaller radii the frequency of the the emitted radiation can be increased further according to $\omega\sim\omega_0\rho_0/\rho$, albeit with deceasing intensities.  This follows from the conservation  of the angular momentum.\\
For comparison, {panel (e) of Fig.\,\ref{fig:fig4}  shows the emission spectrum when applying   a conventional linear polarized Gaussian beam   focused on the center of the multiple-ring complex (meaning roughly the  same waist as for the vortex beam). All rings are  then simultaneously irradiated.
For a credible  comparison with the results for an optical vortex, the intensity is normalized such that both beams have the same number of photons. The photon energy of the two-cycle pulse is 2.5\,meV which is in the regime of the characteristic frequency of the largest ring with 150\,nm. Therefore, it is not surprising to see one strong signal at $\omega/2\pi=0.55$\,THz due to the transitions indicated  in Fig.\,\ref{fig:fig1}(a) (the case for $m_{_{\rm OAM}}=0$). The smaller rings with radii of 100\,nm and 65\,nm are subject to higher intensities due to the Gaussian shaped profile. As a consequence, we find emission signals at higher frequencies despite that the photon energy of the laser pulse does not match the corresponding characteristic frequencies of the rings. This can be explained by multiphoton transitions which, in return, have generally a lower excitation probability \cite{watzel2017ultrafast}. They are not a consequence of tunneling effects because the time delays between the different harmonics is much smaller in comparison to the temporal characteristics shown in the panels (a)-(d) of Fig.\,\ref{fig:fig4}. In contrast to the radiation following  excitation by an optical vortex pulse, all emission signals show a pronounced linear polarization as inferred from the ring-resolved Stokes parameter $\overline{S^z_i}=S^z_i/S^z_0$ for $\omega/2\pi=0.55$\,THz.}

\subsection{Low-frequency emission}

\begin{figure}[t]
\centering
\includegraphics[width=12cm]{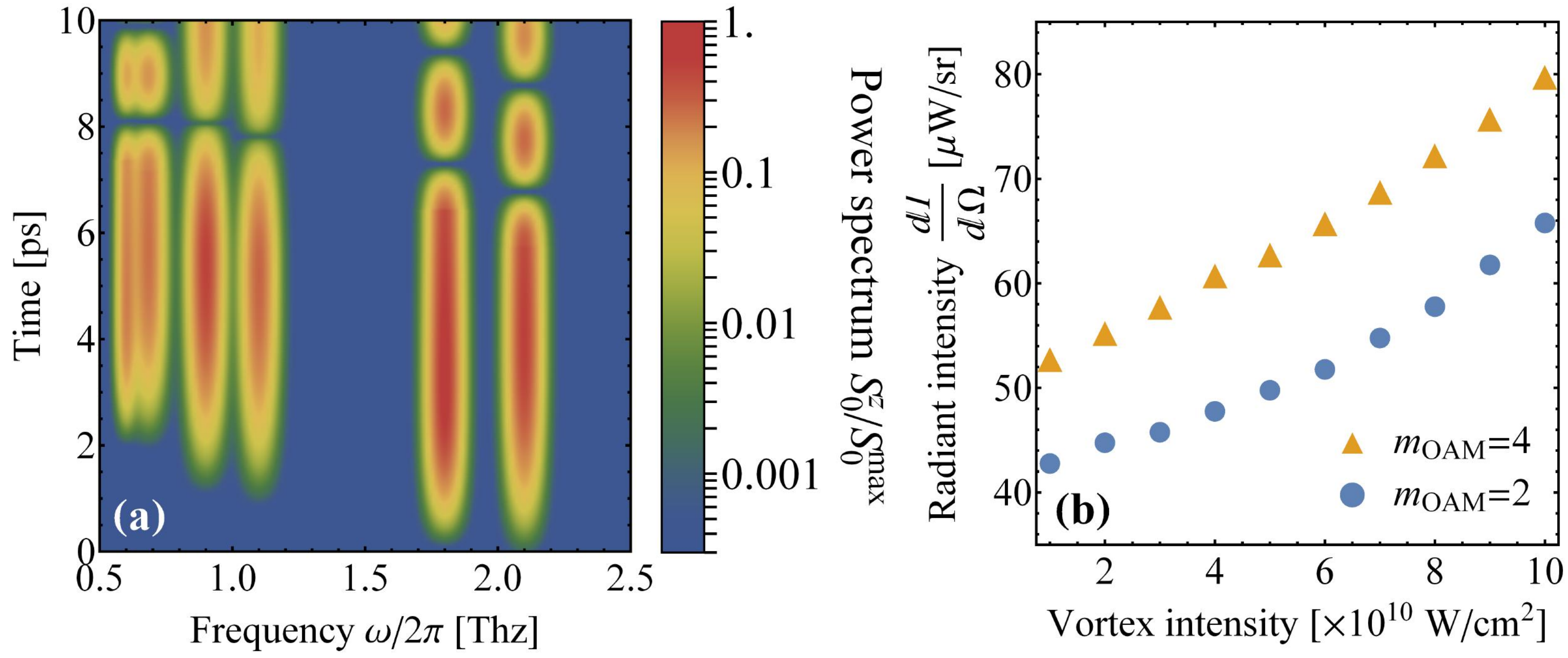}
\caption{(a) Total emission spectrum of the multiple-ring structure. The smallest ring is irradiated by the optical vortex pulse with $m_{_{\rm OAM}}=4$. (b) The emission signal at the converged frequency $\omega/2\pi=0.6$\,THz and a propagation time of 9\,ps in dependence on the intensity of the applied optical vortex pulse.}
\label{fig:fig5}
\end{figure}

The driving frequency can also be  down-converted by simply changing the waist of the vortex beam.
In Fig.\,\ref{fig:fig5}(a) the total time-dependent emission spectrum of the ring-structure  is depicted where, now, the smallest ring was irradiated. The intensity of the applied pulse is unaltered.
The spectrum exhibits  two intense main signals at $\omega/2\pi=1.8$\,THz and $\omega/2\pi=2.1$\,THz which are caused  by resonant-transitions within the ring with the smallest radius $\rho_1=65$\,nm. Through subsequent tunneling processes into the attached rings with larger radii $\rho_2=100$\,nm and $\rho_3=150$\,nm, we find additional emerging signals at lower frequencies. Consequently, the round-trip frequencies are down-converted which is reflected in the characteristics of the emitted radiations.
Inspecting   the times when the additional signals occur and  comparing with the up-conversion spectrum in Fig.\,\ref{fig:fig4} it is obvious that the down-conversion mechanism is faster. Furthermore, the down-conversion process appears more effective as signified by the higher intensity of the additional signals in comparison with fig.\,\ref{fig:fig4}.  This effect stems from the enlarged effective centrifugal potential which leads to an imbalance and a pronounced drift of the charge density to outer ring radii \cite{watzel2016centrifugal}.\\
The centrifugal force experienced by the charge carriers is  stronger for higher    topological charges $m_{_{\rm OAM}}$, as
demonstrated by Fig.\,\ref{fig:fig5}(b) where the down-converted signal at $\omega/2\pi=0.6$\,THz and a propagation time $t=8$\,ps is shown as function  of  the intensity of the applied laser pulse and the topological charge. A higher winding number leads to more intense emission signal after the down-conversion. Furthermore, we deduce a nearly linear dependence of the emission signal on the intensity of the applied vortex pulse, due to the enlarged population of the resonantly excited, current-carrying charge carrier  above the Fermi energy  that   tunnel and trigger loop currents in neighboring rings.

\section{Conclusions}

We demonstrated the dependence of the emission spectrum of an engineered  quantum ring structure on the topological charge of a focused optical vortex driving pulse. In general, the frequency of the radiation can be enlarged for higher vortex winding number, as long as the spectral width of the pulse covers  the relevant optical transitions whose selection rules depend strongly on the winding number $m_{_{\rm OAM}}$.  For one quantum ring   changing the  radiation characteristics  is  limited.
We worked out a possibility to enlarge noticeably the frequency of the emitted harmonics using engineered intercalated rings with exceedingly smaller radii.
Tunneling  orbital currents  between rings allow for the emission of trains of high harmonics with tunable frequency and time structure.  The conversion mechanism can be exploited down to nm size rings, albeit at the cost of strongly decreasing intensities of the higher harmonics.  Low-frequency generation is also possible by simply changing the focusing of the vortex beam.  To achieve yet higher harmonics one may use cut rings, as in  \cite{hinsche2009high}, or side decorated rings which would introduce even  higher frequencies in the emission spectrum.   In principle, one can also just increase the intensity of the vortex beam  in which case the practically  dark spot along the optical axis shrinks and the smaller rings are then excited. This goes however on the cost of having huge peak intensity at larger distances leading to material damage and higher order processes.
\begin{figure}[t]
\centering
\includegraphics[width=12cm]{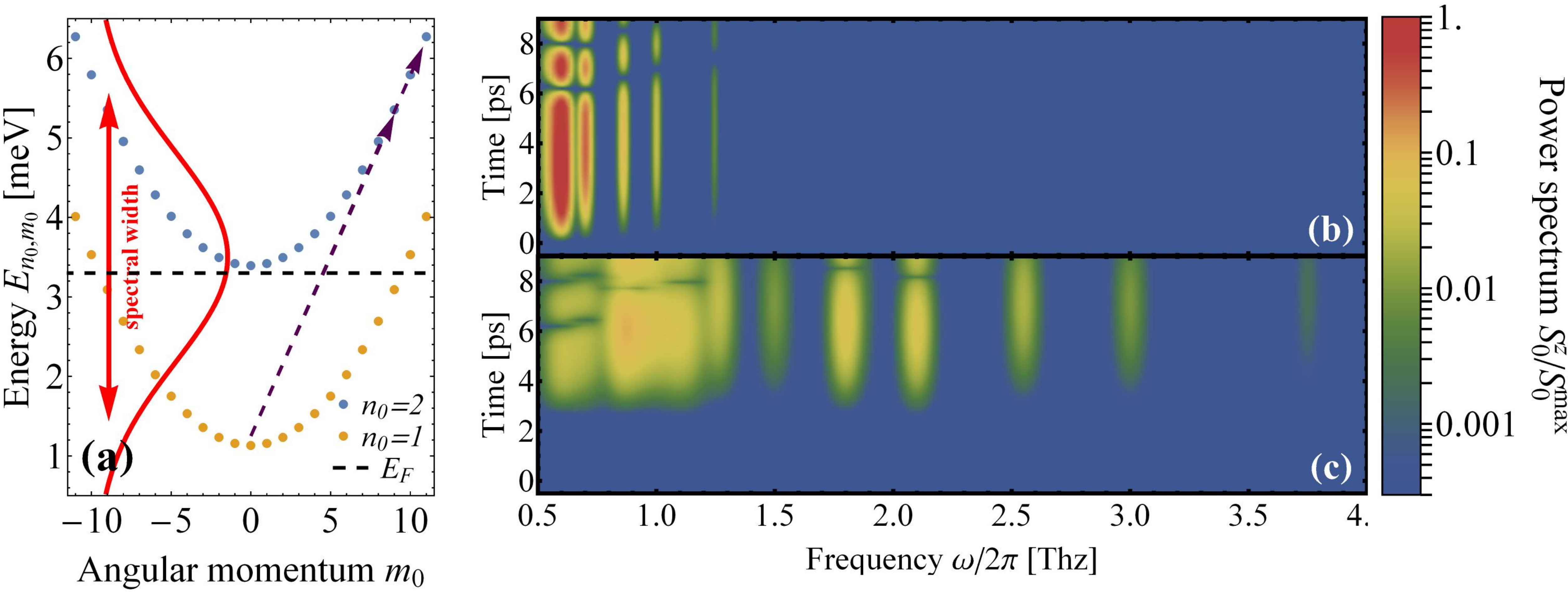}
\caption{(a) Characteristic electronic spectrum for the largest ring of the whole multiple-ring structure  with $\rho_1=150\,$nm. The red curve illustrates the irradiating optical vortex beam with a duration of 1.5 optical cycles and a topological charge $m_{\rm OAM}=10$. (b) Specific time-dependent emission spectrum of the largest ring. (c) Specific time-dependent emission spectrum of the smallest ring with $\rho_1=50\,$nm.}
\label{fig:fig6}
\end{figure}

\section{Appendix}
\subsection{Dependence on the pulse length}

Let us consider the emission characteristics for a multiple-ring structure  driven by a "perfect" optical vortex pulse with a pulse length corresponding to 1.5 optical cycles. Other pulse parameters remain unaltered, i.e. $\hbar\omega_x=2.5$\,eV while the peak intensity is $10^{10}$\,W/cm$^2$. The beam is focussed on the largest ring with $\rho_1=150$\,nm and a width $\Delta\rho_1=40$\,nm. The spatial parameters are $\rho_r=150$\,nm and $w_0=10$\,nm (cf. Eq.\,\eqref{eq:perfectOAM}). A direct consequence of shortening the pulse length is a larger spectral width around the central frequency. In contrast to the situation depicted in Fig.\,\ref{fig:fig1}(a) such a pulse covers electronic states up to quantum numbers $m_0=\pm10$ while the Fourier coefficients of the photon absorption process $A^-=\int_{-\infty}^{\infty}\Omega(t)\exp\left[i((E_{n_{_0}',m_{_0}'}-E_{n_{_0},m_{_0}})/\hbar-\omega_x)t\right]$ are still much larger than the Fourier coefficients of the photon emission process $A^+=\int_{-\infty}^{\infty}\Omega(t)\exp\left[i((E_{n_{_0}',m_{_0}'}-E_{n_{_0},m_{_0}})/\hbar+\omega_x)t\right]$.
A schematics   is shown  in Fig.\,\ref{fig:fig6}(a). In panel (b) of Fig.\,\ref{fig:fig6} the time-dependent emission spectrum for the largest ring is depicted, meaning that  the integration domain in Eq.\,\eqref{eq:dipx} and Eq.\,\eqref{eq:dipy} is restricted to $[\rho_1-\Delta\rho_1/2,\rho_1+\Delta\rho_1/2]$. The topological charge of the applied optical vortex pulse is $m_{_{\rm OAM}}=10$. In comparison to the results displayed  in Fig.\,\ref{fig:fig2}(b) or Fig.\,\ref{fig:fig4}(a) we identify  additional signals at higher frequencies  due to optical selection rules and the large spectral width of the pulse. Several electronic states are excited and contribute to the radiation of the quantum ring. The values of the different harmonics follow from the energy differences between the charge carrier states in the characteristic electronic structure.
Similarly.  the frequencies of the radiation  can be up-converted  by  tunneling processes as shown in panel (c) of Fig.\,\ref{fig:fig6}.  The signals appear time-delayed in the smaller rings. The results illustrate the strong dependence of the whole up-conversion process on the spectral width of the applied optical vortex pulse which, in turn, determines the maximal usable topological charge $m_{_{\rm OAM}}$ and therefore the highest achievable value of the converted harmonics. The Stokes parameter (not depicted here) corresponding to the different harmonics indicate a pronounced circular polarization ($\overline{S^z_3}\approx1$).

\bibliographystyle{osajnl}

\end{document}